\documentclass[letterpaper,10pt]{article} 

\usepackage{osameet3} 

\usepackage{amsmath,amssymb}
\usepackage[colorlinks=true,bookmarks=false,citecolor=blue,urlcolor=blue]{hyperref} 

\begin{document}

\title{High-sensitivity frequency modulation CARS with a compact and fast tunable fiber-based light source}

\author{Thomas Würthwein$^{1,*}$, Maximilian Brinkmann$^{2}$, Tim Hellwig$^{2}$, Kristin Wallmeier$^{1}$, and Carsten Fallnich$^{1,3,4}$}
\address{$^1$ Institute of Applied Physics, University of Münster, Corrensstraße 2, 48149 Münster, Germany \\ 
$^2$ Refined Laser Systems GmbH, Mendelstraße 11, 48149 Münster, Germany \\
$^3$ MESA+ Institute for Nanotechnology, University of Twente, P.O. Box 217, Enschede 7500 AE, The Netherlands \\
$^4$ Cells in Motion Interfaculty Centre, University of Münster, Waldeyerstraße 15, 48149 Münster, Germany \\}
\email{*t.wuerthwein@uni-muenster.de}

\copyrightyear{2021}

\begin{abstract}
Frequency modulation coherent anti-Stokes Raman scattering (FM CARS) is presented, using a compact as well as fast and widely tunable fiber-based light source. With this light source Raman resonances  between 700\,cm$^{-1}$ and 3200\,cm$^{-1}$ can be addressed via wavelength tuning within only 5\,ms, which allows for FM CARS measurements with frame-to-frame wavelength switching. Moreover, the functionality for high-sensitivity FM CARS measurements was integrated by means of fiber optics to keep a stable and reliable operation. The light source accomplished FM CARS measurements with a 40 times enhanced sensitivity at a lock-in amplifier (LIA) bandwidth of 1\,Hz. For fast imaging with frame-to-frame wavelength switching at a LIA bandwidth of 1\,MHz an 18-fold contrast enhancement could be verified, making this light source ideal for routine and out-of-lab FM CARS measurements for medical diagnostics or environmental sensing.
\end{abstract}



\maketitle

\section{Introduction}
Coherent anti-Stokes Raman scattering (CARS) microscopy has become a powerful technique with a number of applications in the fields of biomedical imaging, cell biology and medicine \cite{Evans2008}. If a pump and a Stokes field, with the frequencies $\omega_\text{P}$ and $\omega_\text{S}$, respectively, interact with a Raman-active molecular vibrational resonance at the frequency $\Omega = \omega_\text{P} - \omega_\text{S}$ a resonant anti-Stokes signal at the frequency $\omega_{\text{AS}} = 2 \omega_\text{P} - \omega_\text{S}$ is generated. This signal allows for chemically-selective imaging of unstained samples.

However, with the resonant signal there comes also a nonresonant contribution, which does not contain any specific chemical information. This nonresonant background distorts the resonant signal and, depending on the sample, can even overwhelm the resonant signal \cite{Lotem1976}. The resonant and nonresonant CARS responses originate from the third-order susceptibility and were studied in detail by Lotem et. al \cite{Lotem1976}.

The detection of the CARS signal in epi-direction significantly reduces the nonresonant contribution and consequently improves the sensitivity \cite{Cheng2001a}. Nevertheless, a number of alternative techniques to avoid or eliminate the nonresonant background in CARS were presented, e.g., polarization-sensitive detection \cite{Nestor1978, Oudar1979a, Cheng2001, Wurthwein2017}, and time-resolved CARS \cite{Kamga1980, Volkmer2002}, which also come along with an attenuated signal leading to longer acquisition times. Broadband techniques, such as multiplex CARS (M-CARS), allow for the reconstruction of the original Raman lineshape \cite{Wurpel2002}, with the disadvantage of long integration times not suitable for high-speed imaging applications. Interferometric CARS provides sufficient imaging speed and sensitivity \cite{Oron2002,Potma2006}, but suffers from image artifacts for samples with a variable index of refraction. Furthermore, the digital subtraction of resonant and nonresonant images was presented and allowed for the acquisition of background-corrected images \cite{Li2016}. As an alternative technique for the acquisition of background-corrected CARS signals, frequency modulation (FM) CARS was demonstrated theoretically \cite{Lotem1983} and experimentally \cite{Ganikhanov2006b,Chen2010}. In FM CARS the resonant as well as the nonresonant contribution of the CARS signal are measured by two wavelength-alternating pump pulses together with a Stokes pulse fixed in wavelength. Lock-in amplifier (LIA) detection is then used for immediate difference calculation between the resonant and the nonresonant CARS signal at the switching frequency of the two alternating pump waves. Accordingly, FM CARS allows for high-speed acquisition of background-corrected CARS signals with an enhanced sensitivity.

The first experimental realizations of FM CARS  based on the combination of different solid-state light sources providing the frequency-modulated pump fields and the Stokes field. Nevertheless, with the combination of these light sources concentration values down to 0.05\,\% could be measured, which was approximately two orders of magnitude better than in standard CARS \cite{Ganikhanov2006b}. Later, FM CARS using chirped laser pulses from a single Ti:sapphire laser was presented,  which already reduced the complexity of the light source, but was limited in terms of the tuning speed and the tuning range either in the CH-stretch \cite{Rocha-Mendoza2009} or the fingerprint region \cite{Chen2010}. However, the complexity and high demand for maintenance of free-space light sources for FM CARS did not allow routine application outside a specialised laser laboratory.

In order to overcome the above mentioned limitations, we realized a compact as well as fast and widely tunable fiber-based light source providing all necessary pulses for FM CARS. Using this light source, Raman resonances between 700\,cm$^{-1}$ and 3200\,cm$^{-1}$ were addressable via wavelength tuning within only 5\,ms for an arbitrary wavelength step, and enabled high-sensitivity FM CARS measurements with frame-to-frame wavelength switching. 
This FM functionality is based on fiber optics and, therefore, seamlessly integrateable into an all-fiber FOPO light source \cite{Brinkmann2019}. We present concentration measurements using FM CARS achieving a higher sensitivity compared to standard CARS microscopy. Moreover, FM CARS imaging and dual-color FM CARS imaging with an enhanced contrast are shown. 
As the entire light source is based on fiber-optics a compact and robust light source is realized. This development constitutes an important step for advancing coherent Raman imaging in terms of portability and sensitivity for applications in medical diagnostics or environmental sensing.

\begin{figure}[htbp]
\centering
\includegraphics[width = 0.75\linewidth]{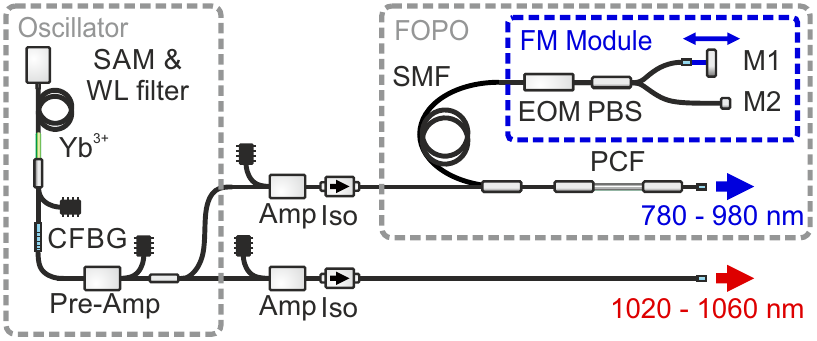}
\caption{Scheme of the fiber-based light source for FM CARS with the integrated module for frequency modulation (FM module, blue box). For details see text.}
\label{fig1:setup}
\end{figure}

An all-fiber light source consisting of an ytterbium-doped (Yb$^{3+}$) fiber oscillator and a FOPO was realized (Fig.~\ref{fig1:setup}), which delivered synchronized picosecond pulses tunable in wavelength for coherent Raman measurements. All specifications presented in Ref.~\cite{Brinkmann2019} hold for the here presented modified version of the light source, which provided three instead of two pulses at different wavelengths, namely the Stokes pulse as well as the wavelength-alternating pump pulses, necessary for the later FM CARS experiments. The Stokes pulses, generated within the oscillator and mode-locked with a saturable absorber mirror (SAM), were electronically tunable in wavelength (WL filter) between 1020\,nm and 1060\,nm, had a duration of 7\,ps, and were amplified (Pre-Amp and Amp) in two stages to 400\,mW at the output behind an optical isolator (Iso). The wavelength filter consisted of a custom-made fiber-coupled filter based on an optical grating and a compact electro-mechanical beam deflection with a switching time of 300\,µs, which is much faster than the FOPO startup of approximately 5\,ms. 
The beam quality factor was M$^2=1.16 \pm 0.07$ and a relative intensity noise (RIN) at 20.25\,MHz of -153.5\,dBc/Hz was measured. Beside the Stokes pulses, 7\,ps long pump pulses were generated in a FOPO with a linear resonator, which was based on 50\,cm of polarization-maintaining (PM) photonic-crystal fiber (PCF, NKT Photonics, LMA-PM-5) and about 155\,m of PM single-mode fiber (SMF, Nufern, PM780-HP) between a custom-made FM module and a polished FC/PC connector at the output with a reflectivity of about 4\,\%. The PCF was used to generate parametric four-wave mixing (FWM) gain tunable in wavelength between 750\,nm and 980\,nm by wavelength tuning of the oscillator within only 5\,ms (the corresponding wavelength tuning curve can be found in Fig.~3(a) of Ref. \cite{Brinkmann2019}). The combination of the FOPO and the amplified oscillator pulses served as the pump and Stokes waves for CARS, and allowed to address Raman bands between 700\,cm$^{-1}$ and 3200\,cm$^{-1}$. The SMF in the FOPO resonator accomplished a spectrally narrow dispersive tuning \cite{Yamashita2006, Brinkmann2016}, such that the feedback signal pulses were stretched in time, and only a narrow spectral part ($<12\,$cm$^{-1}$) overlapped with the next pump pulse to be amplified. Therefore, the optical path length of the resonator was directly connected to the wavelength of the FOPO output. The custom-made chirped fiber Bragg grating (CFBG) between the FOPO and the oscillator was used to match the repetition rate of the oscillator to the repetition rate of the FOPO for all oscillator wavelengths and replaced a free-space optical delay line as used in other FOPOs \cite{Gottschall2015} to keep oscillator and FOPO synchronized. For the output of the FOPO a beam quality factor of M$^2=1.03 \pm 0.03$ and a RIN of -127.5\,dBc/Hz was measured.

\begin{figure}[htbp]
\centering
\includegraphics[width = 0.75\linewidth]{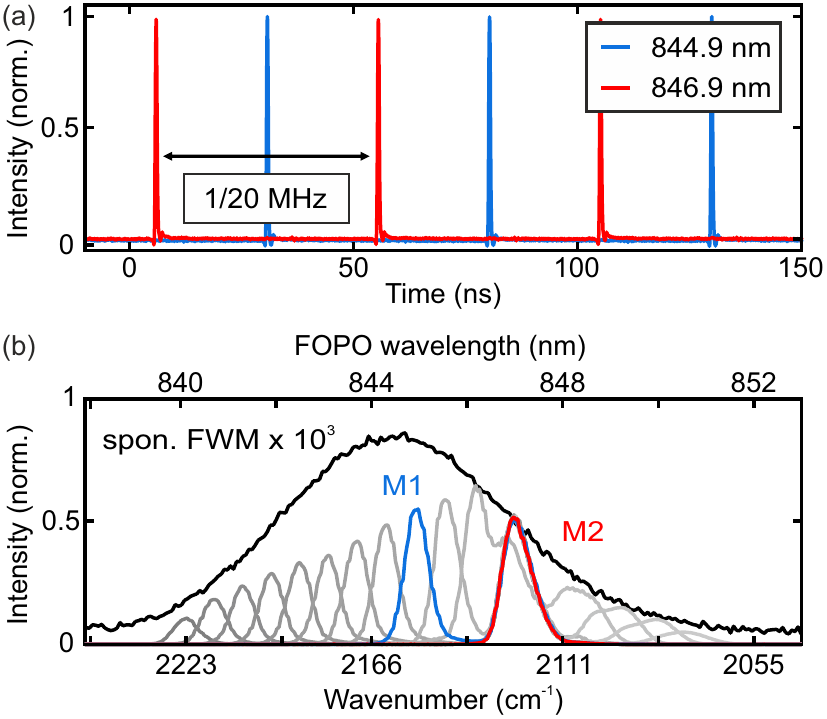}
\caption{(a) Temporal output of the FOPO measured with two photo-detectors, verifying the pulse-to-pulse wavelength alternation. (b) Wavelength tunability within the spontaneous FWM gain (black curve) by the feedback of mirror M1 (blue) and M2 (red curve). Grey curves represent the FOPO output for different positions of M1.}
\label{fig2:specs}
\end{figure}

For the FM CARS measurements the feedback mechanism was modified as highlighted in the blue box (FM module, Fig.~\ref{fig1:setup}) to form in principle two optically separated FOPO resonators, where every second pulse was fed back through a different path. For this purpose the fiber-coupled end-mirror of the FOPO was replaced within the conventional setup of Ref.~\cite{Brinkmann2019} by a module consisting of a fiber-coupled electro-optic modulator (EOM), a fiber-coupled polarizing beam splitter (PBS) and two end-mirrors (M1 and M2). The EOM was synchronized to half of the 40.5\,MHz oscillator repetition rate, which led to a pulse-to-pulse switching between the two mirrors M1 and M2, respectively. As the optical path length between the PBS and M1 was different compared to that between the PBS and M2, two linear resonators with different optical path lengths were formed, which resulted due to the dispersive tuning in two alternating center wavelengths of the output pulses of the FOPO. 

The pulse-to-pulse wavelength switching of the FOPO is exemplarily shown for a fixed Stokes wavelength of 1032.7\,nm (Fig.~\ref{fig2:specs}(a)). The wavelength switching between 844.9\,nm (2152\,cm$^{-1}$) and 846.9\,nm (2124\,cm$^{-1}$) was investigated by separating the FOPO output in wavelength by means of a grating and measuring the two pulse trains separated in space with two photo-detectors. The resulting time traces verify the clear pulse-to-pulse alternation in wavelength.  

Beside the temporal profile, the spectral output of the FOPO around 845\,nm was measured using an optical spectrum analyzer (Figs.~\ref{fig2:specs}(b)), while the Stokes pulses stayed centered at 1032.7\,nm wavelength. This wavelength combination gave access to the  spectral region of deuterated samples, e.g. deuterated dimethyl sulfoxide (dDMSO). For wavelength fine-tuning within the spontaneous FWM gain region (black curve in Fig.~\ref{fig2:specs}(b)) the optical path length difference of the feedback provided by the mirrors M1 and M2 was adjusted. The spectra highlighted in red and blue are separated by 28\,cm$^{-1}$ and represent the measured pulse trains from Fig.~\ref{fig2:specs}(a). In this configuration, mirror M2 was fiber-integrated and fixed, accomplishing the FOPO to emit at a fixed wavenumber on the right side of the gain bandwidth (red curve) for a fixed oscillator wavelength. Whereas, M1 was placed on a free-space optical delay line to precisely adjust via its position the second wavelength in the FWM gain region (grey curves). To cover the entire spontaneous FWM gain bandwidth the mirror M1 had to be moved by 5.6\,cm, whereas the difference in the resonator length for the red and blue spectra in Fig.~2(b) was 0.8\,cm only and stayed fixed during the measurement. As a consequence, if pulses with a certain and fixed wavelength difference at the output are acceptable the free-space optical delay line could be replaced by a completely fiber-integrated mirror.


All CARS measurements were performed in forward-direction using a home-built laser-scanning microscope with a microscope objective (Seiwa PEIR-Plan-50x, NA = 0.6) and a photo-multiplier-tube (PMT, Hamamatsu H7422-20) without descanning. The PMT signal was analyzed with a LIA (Zurich Instruments HF2LI) at the modulation frequency of 20.25\,MHz. For the FM CARS measurements the FOPO as shown in Fig.~\ref{fig1:setup} was used, whereas for standard CARS measurements the feedback path of M1 was blocked by a mechanical shutter. 

 \begin{figure}[htbp]
\centering
\includegraphics[width = 0.75\linewidth]{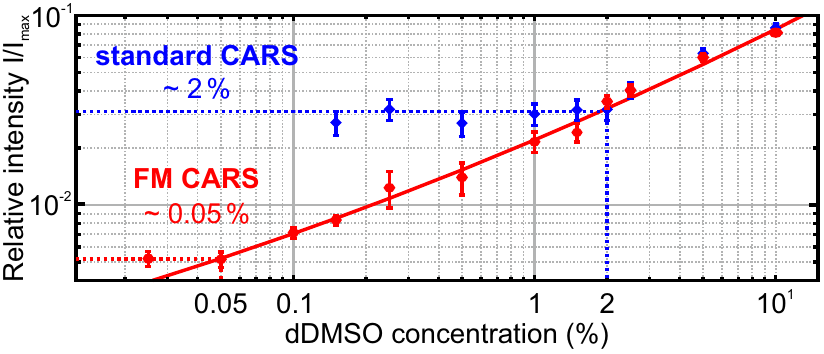}
\caption{Dilution series of dDMSO in water, where $I$ is the signal from dDMSO dissolved in water normalized to the signal $I_{\text{max}}$ of pure dDMSO. The standard CARS signal is shown as blue diamonds, whereas the FM CARS signal is visualized as red dots. Straight dotted lines were added to guide the eye, and the red solid line represents a fit of Eq.~(2) from Ref. \cite{Ganikhanov2006b}.}
\label{fig:dilution_series}
\end{figure}
In order to quantify the achieved increase in detection sensitivity of FM CARS in comparison to standard CARS a dilution series with dDMSO and water was measured. For this measurement the resonance of dDMSO at 2125\,cm$^{-1}$ and the nonresonant contribution at about 2145\,cm$^{-1}$ were addressed with the same average power of about 20\,mW in the imaging plane.  For this specific measurement, the LIA detection bandwidth was set to 1\,Hz for reducing the noise to enhance the sensitivity. 
 
 The standard CARS signal as well as the FM CARS signal were each normalized to the signal of pure dDMSO and plotted against concentration (Fig.~\ref{fig:dilution_series}). In order to verify the benefits of FM CARS we investigated low concentrations of less than $10\,\%$. With standard CARS a concentration of approximately 2\,\% dDMSO in water could be measured, limited by the nonresonant background. For FM CARS the detection limit was approximately at 0.05\,\% dDMSO in water (red and blue dotted lines were added to guide the eye) and thus the sensitivity was improved by a factor of 40. Even lower concentrations could not be resolved due to residual electronic noise. The FM CARS results were in good agreement with Eq.~(2) (red solid line) from Ref. \cite{Ganikhanov2006b}, in which  Ganikhanov et al. introduced an expression for the FM CARS intensity at low concentrations.

 The measured concentration limit using FM CARS of approximately 0.05\,\% dDMSO in water is as good as it was presented in Ref. \cite{Ganikhanov2006b} for methanol in water. However, instead of using synchronized solid-state lasers, the here presented results were achieved by using a single compact and robust light source, which is a significant simplification and enables the routine application of FM CARS in, but also outside specialized laser laboratories.

\begin{figure}[htbp]
\centering
\includegraphics[width = 0.75\linewidth]{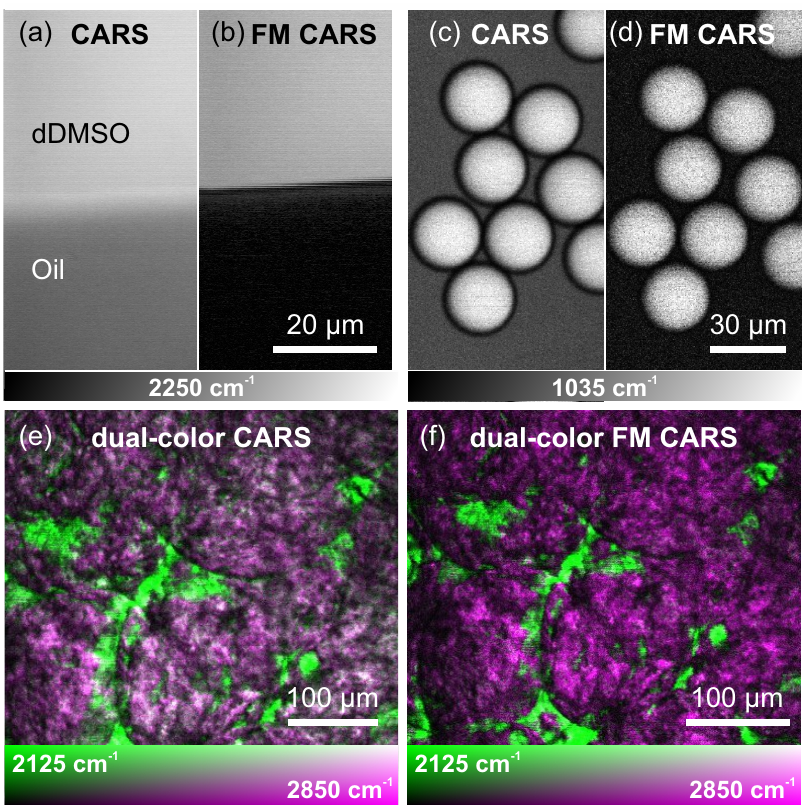}
\caption{Comparison between CARS and FM CARS images: (a,\,b) interface between dDMSO and rapeseed oil measured at 2250\,cm$^{-1}$, (c, d) polystyrene beads in rapeseed oil addressing the resonance at 1035\,cm$^{-1}$ in the fingerprint region, and (e, f) merged dual-color images of adipose tissue soaked with dDMSO, addressing the Raman resonances of dDMSO and lipids at 2125\,cm$^{-1}$ and 2850\,cm$^{-1}$, respectively. The second pump wavelength, addressing the nonresonant CARS contribution in FM CARS, was set approx. 20\,cm$^{-1}$ above the mentioned wavenumbers.}
\label{fig:contrast_enhancement}
\end{figure}

In order to show the capability of the presented light source for imaging applications, at first a technical sample, i.e. a mixture of dDMSO and rapeseed oil was imaged using standard CARS and FM CARS for direct comparison (Fig.~\ref{fig:contrast_enhancement}(a) and (b)). The wavelengths of the modulated pump pulses were 841.4\,nm and 840.5\,nm, addressing in combination with the Stokes pulses at 1038\,nm the weak Raman resonance of dDMSO at 2250\,cm$^{-1}$ and the nonresonant contribution at 2263\,cm$^{-1}$, respectively. The investigated rapeseed oil only showed a nonresonant CARS signal, as rapeseed oil does not exhibit any Raman resonance at both wavenumbers. In the focal plane, the two pump pulse trains had each an average power of 32\,mW, whereas the Stokes pulse train had an average power of 60\,mW. All images (512x512 pixels, no averaging) shown in Fig.~\ref{fig:contrast_enhancement} were acquired with a LIA bandwidth of 1\,MHz to demonstrate that the frequency-modulated FOPO is well suited for fast imaging, although a lower LIA bandwidth would further reduce the electronic noise. As visualized in Fig.~\ref{fig:contrast_enhancement}(a), a strong nonresonant CARS signal from the rapeseed oil (lower part) was measurable, whereas in Fig.~\ref{fig:contrast_enhancement}(b) the FM CARS signal in the region of the rapeseed oil showed a nonresonant signal near zero and, therefore, a higher contrast. The contrast was quantified by dividing the averages of the nonresonant and the resonant signal from the images in Fig.~\ref{fig:contrast_enhancement}(a) and (b) and was in the standard CARS image 2.5, whereas 45.0 in FM CARS. Thus, an 18-fold contrast enhancement and a significantly improved distinguishability were achieved.

In order to demonstrate the FM CARS functionality of acquiring background-suppressed images addressing different resonances in the Raman spectrum, polystyrene (PS) beads in rapeseed oil (Fig.~\ref{fig:contrast_enhancement}(c) and (d)) and fresh adipose tissue soaked for two hours with dDMSO (Fig.~\ref{fig:contrast_enhancement}(e) and (f)) were imaged. The PS beads were imaged addressing the fingerprint resonance at 1035\,cm$^{-1}$. In this particular sample, a contrast enhancement by a factor of 7.2 was determined. In general, the achievable contrast enhancement depends on the strength of the resonant as well as the nonresonant CARS contributions and differs for different Raman resonances and samples. Standard CARS (Fig.~\ref{fig:contrast_enhancement}(e)) and FM CARS (Fig.~\ref{fig:contrast_enhancement}(f)) images of the tissue were acquired addressing the molecular vibrations of dDMSO at 2125\,cm$^{-1}$ (from black to green on the vertical axis of the 2D colormap) and lipids at 2850\,cm$^{-1}$ (from black to magenta on the horizontal axis of the 2D colormap). In the standard CARS image a strong nonresonant contribution of the lipids reduced the contrast, which led to the whitish regions in Fig.~\ref{fig:contrast_enhancement}(e) as green and magenta add up to white (diagonal of the 2D colormap). FM CARS enabled imaging with a reduced nonresonant contribution of the lipids, and, therefore, the green and magenta areas became spatially well separated in the false-color image in Fig.~\ref{fig:contrast_enhancement}(f). For all FM CARS images the nonresonant CARS contribution was measured at a wavenumber roughly 20\,cm$^{-1}$ above the wavenumber of the mentioned Raman resonances. Beside the images in Fig.~\ref{fig:contrast_enhancement} we attached \textcolor{blue}{Visualization 1} to the supplementary material, which shows the real-time imaging of the dDMSO and lipid resonances in fresh adipose tissue with standard and FM CARS at four frames per second (256x256 pixels, no averaging), however, with frame-to-frame wavelength-switching. Especially, this measurement emphasises the benefits of the here presented light source, as it was possible to switch between two Raman resonances in less than 5\,ms (verified by a measurement shown in Ref. \cite{Brinkmann2019}) combined with the contrast enhancement of FM CARS. 

In conclusion, a robust and compact, as well as fast and widely tunable fiber-based light source for FM CARS was presented. The light source consisted of a fast tunable mode-locked fiber oscillator, a two-stage amplifier and a fiber optical parametric oscillator adapted to FM CARS with a fiber-based module for wavelength alternation at 20.25\,MHz modulation frequency. In combination with the Stokes pulses, background-suppressed FM CARS imaging was accomplished in a simple lock-in amplifier (LIA) detection scheme. The usage of FM CARS allowed for the detection of concentrations as low as 0.05\,\% at a LIA bandwidth of 1\,Hz, representing an improvement in sensitivity by a factor of 40 in comparison to standard CARS. At a LIA bandwidth of 1\,MHz the reduction of the nonresonant contribution of the CARS signal and, therefore, the increased detection sensitivity resulted in a contrast enhancement by a factor of 18, resulting in an improved image quality even for fast imaging applications. Combined with the fast and wide wavelength tunability of the light source within 5\,ms only, FM CARS with frame-to-frame wavelength switching was accomplished. In the future, the FM functionality of the presented light source could also be used for the accurate measurement of the ratio of two adjacent Raman resonances with a time resolution of 50\,ns, by referencing one Raman resonance against a second Raman resonance, which for instance could find applications in the field of tumor diagnostics \cite{Sarri2019}. Moreover, both the compact fiber-integration and the achieved robustness enable FM CARS imaging not only inside a specialized laser-laboratory, but offers the potential to be used for medical diagnostics or environmental sensing. \\

\noindent\textbf{Disclosures.} Maximilian Brinkmann and Tim Hellwig (Refined Laser Systems GmbH) declare a commercial interest. \\


\begin{thebibliography}{10}

\bibitem{Evans2008}
Conor~L. Evans and X.~Sunney Xie.
\newblock {Coherent Anti-Stokes Raman Scattering Microscopy: Chemical Imaging
  for Biology and Medicine}.
\newblock {\em Annu. Rev. Anal. Chem.}, 1(1):883--909, jul 2008.

\bibitem{Lotem1976}
Haim Lotem, R.~T. Lynch, and N.~Bloembergen.
\newblock {Interference between Raman resonances in four-wave difference
  mixing}.
\newblock {\em Phys. Rev. A}, 14(5):1748--1755, nov 1976.

\bibitem{Cheng2001a}
Ji-Xin Cheng, Andreas Volkmer, Lewis~D Book, and X~Sunney Xie.
\newblock {An Epi-Detected Coherent Anti-Stokes Raman Scattering (E-CARS)
  Microscope with High Spectral Resolution and High Sensitivity}.
\newblock {\em J. Phys. Chem. B}, 105(7):1277--1280, feb 2001.

\bibitem{Nestor1978}
J~R Nestor.
\newblock {Polarization properties of coherent anti-stokes Raman spectra (CARS)
  in isotropic liquids}.
\newblock {\em J. Raman Spectrosc.}, 7(2):90--95, apr 1978.

\bibitem{Oudar1979a}
Jean‐Louis Oudar, Robert~W. Smith, and Y.~R. Shen.
\newblock {Polarization‐sensitive coherent anti‐Stokes Raman spectroscopy}.
\newblock {\em Appl. Phys. Lett.}, 34(11):758--760, jun 1979.

\bibitem{Cheng2001}
Ji-Xin Cheng, Lewis~D Book, and X~Sunney Xie.
\newblock {Polarization coherent anti-Stokes Raman scattering microscopy}.
\newblock {\em Opt. Lett.}, 26(17):1341, sep 2001.

\bibitem{Wurthwein2017}
Thomas W{\"{u}}rthwein, Maximilian Brinkmann, Tim Hellwig, and Carsten
  Fallnich.
\newblock {Rapid spectro-polarimetry to probe molecular symmetry in multiplex
  coherent anti-Stokes Raman scattering}.
\newblock {\em J. Chem. Phys.}, 147(19), 2017.

\bibitem{Kamga1980}
Francois~M. Kamga and Mark~G. Sceats.
\newblock {Pulse-sequenced coherent anti-Stokes Raman scattering spectroscopy:
  a method for suppression of the nonresonant background}.
\newblock {\em Opt. Lett.}, 5(3):126, mar 1980.

\bibitem{Volkmer2002}
Andreas Volkmer, Lewis~D. Book, and X.~Sunney Xie.
\newblock {Time-resolved coherent anti-Stokes Raman scattering microscopy:
  Imaging based on Raman free induction decay}.
\newblock {\em Appl. Phys. Lett.}, 80(9):1505--1507, mar 2002.

\bibitem{Wurpel2002}
George W~H Wurpel, Juleon~M Schins, and Michiel M{\"{u}}ller.
\newblock {Chemical specificity in three-dimensional imaging with multiplex
  coherent anti-Stokes Raman scattering microscopy.}
\newblock {\em Opt. Lett.}, 27(13):1093--5, 2002.

\bibitem{Oron2002}
Dan Oron, Nirit Dudovich, and Yaron Silberberg.
\newblock {Single-Pulse Phase-Contrast Nonlinear Raman Spectroscopy}.
\newblock {\em Phys. Rev. Lett.}, 89(27):1--4, 2002.

\bibitem{Potma2006}
Eric~O Potma, Conor~L Evans, and X~Sunney Xie.
\newblock {Heterodyne coherent anti-Stokes Raman scattering (CARS) imaging}.
\newblock {\em Opt. Lett.}, 31(2):241, jan 2006.

\bibitem{Lotem1983}
Haim Lotem.
\newblock {Frequency modulation coherent anti-Stokes Raman spectroscopy
  (FM-CARS): A novel sensitive nonlinear optical method}.
\newblock {\em J. Chem. Phys.}, 79(5):2177--2180, 1983.

\bibitem{Ganikhanov2006b}
Feruz Ganikhanov, Conor~L. Evans, Brian~G. Saar, and X.~Sunney Xie.
\newblock {High-sensitivity vibrational imaging with frequency modulation
  coherent anti-Stokes Raman scattering (FM CARS) microscopy}.
\newblock {\em Opt. Lett.}, 31(12):1872, 2006.

\bibitem{Chen2010}
Bi-Chang Chen, Jiha Sung, and Sang-Hyun Lim.
\newblock {Chemical Imaging with Frequency Modulation Coherent Anti-Stokes
  Raman Scattering Microscopy at the Vibrational Fingerprint Region}.
\newblock {\em J. Phys. Chem. B}, 114(50):16871--16880, dec 2010.

\bibitem{Rocha-Mendoza2009}
Israel Rocha-Mendoza, Wolfgang Langbein, Peter Watson, and Paola Borri.
\newblock {Differential coherent anti-Stokes Raman scattering microscopy with
  linearly chirped femtosecond laser pulses}.
\newblock {\em Opt. Lett.}, 34(15):2258, aug 2009.

\bibitem{Chen2010a}
Bi-Chang Chen, Jiha Sung, and Sang-Hyun Lim.
\newblock {Frequency modulation coherent anti-Stokes Raman scattering (FM-CARS)
  microscopy based on spectral focusing of chirped laser pulses}.
\newblock {\em Multiphot. Microsc. Biomed. Sci. X}, 7569:756909, 2010.

\bibitem{Brinkmann2019}
Maximilian Brinkmann, Alexander Fast, Tim Hellwig, Isaac Pence, Conor~L. Evans,
  and Carsten Fallnich.
\newblock {Portable all-fiber dual-output widely tunable light source for
  coherent Raman imaging}.
\newblock {\em Biomed. Opt. Express}, 10(9):4437, sep 2019.

\bibitem{Yamashita2006}
Shinji Yamashita and Masahiro Asano.
\newblock {Wide and fast wavelength-tunable mode-locked fiber laser based on
  dispersion tuning}.
\newblock {\em Opt. Express}, 14(20):9399, 2006.

\bibitem{Brinkmann2016}
Maximilian Brinkmann, Sarah Janfr{\"{u}}chte, Tim Hellwig, Sven Dobner, and
  Carsten Fallnich.
\newblock {Electronically and rapidly tunable fiber-integrable optical
  parametric oscillator for nonlinear microscopy}.
\newblock {\em Opt. Lett.}, 41(10):2193, may 2016.

\bibitem{Gottschall2015}
Thomas Gottschall, Tobias Meyer, Martin Baumgartl, Cesar Jauregui, Michael
  Schmitt, J{\"{u}}rgen Popp, Jens Limpert, and Andreas T{\"{u}}nnermann.
\newblock {Fiber-based light sources for biomedical applications of coherent
  anti-Stokes Raman scattering microscopy}.
\newblock {\em Laser Photon. Rev.}, 9(5):435--451, sep 2015.

\bibitem{Sarri2019}
Barbara Sarri, Rafa{\"{e}}l Canonge, Xavier Audier, Emma Simon, Julien Wojak,
  Fabrice Caillol, C{\'{e}}cile Cador, Didier Marguet, Flora Poizat, Marc
  Giovannini, and Herv{\'{e}} Rigneault.
\newblock {Fast stimulated Raman and second harmonic generation imaging for
  intraoperative gastro-intestinal cancer detection}.
\newblock {\em Sci. Rep.}, 9(1):1--10, 2019.

\end{thebibliography}
\end{document}